 \documentclass[final,1p,times]{elsarticle}

\usepackage{amssymb}
\usepackage{amsmath}
\usepackage{subfigure}
\usepackage{multirow}
\usepackage{bm}
\usepackage{color}

\biboptions{sort&compress}

\newcommand{\vett}[1]{\boldsymbol{#1}}

\journal{Physics Letters B}

\begin{document}

\begin{frontmatter}



\title{Hadronization Scheme Dependence of Long-Range Azimuthal Harmonics
in High Energy p+A Reactions}


\author[col]{Angelo Esposito\corref{cor1}}
\ead{aesposito2458@columbia.edu}
\cortext[cor1]{Corresponding author}

\author[col]{Miklos Gyulassy}

\address[col]{Department of Physics, Columbia University, 538 West 120th Street, New York, NY 10027, USA}

\begin{abstract}
We compare the distortion effects of three popular final-state 
hadronization schemes. We show how hadronization modifies the initial-state gluon correlations 
in high energy p+A collisions. The three models considered are (1) LPH: local parton-hadron duality,
(2) CPR: collinear parton-hadron resonance independent fragmentation, and (3) LUND: color string hadronization. The strong 
initial-state azimuthal asymmetries are generated using the GLVB model for non-abelian gluon bremsstrahlung, assuming a saturation scale $Q_\text{sat}=2$ GeV. Long-range elliptic and triangular harmonics for the final hadron pairs  
are compared based on the three hadronization schemes.
Our analysis shows that the process of hadronization causes major distortions of the partonic azimuthal harmonics for transverse momenta at least up to $p_T=3$ GeV. In particular, they appear to be greatly reduced for $p_T<1\div2$ GeV.
\end{abstract}


\end{frontmatter}


\section{Introduction and motivation}
\label{sec:intro}
Multi-particle, long-range in pseudo-rapidity  azimuthal correlations  are widely studied in high-energy nuclear collisions at RHIC
Au+Au~\cite{STARAA,PHENIXAA} and LHC Pb+Pb~\cite{ALICEAA,CMSAA,ATLASAA}. In particular, they are considered as a signature for the ``perfect fluid'' behavior 
of the strongly coupled Quark-Gluon-Plasma (sQGP) produced in such reactions.

\begin{figure}[t]
\centering
\subfigure[]{
\includegraphics[width=0.46\textwidth]{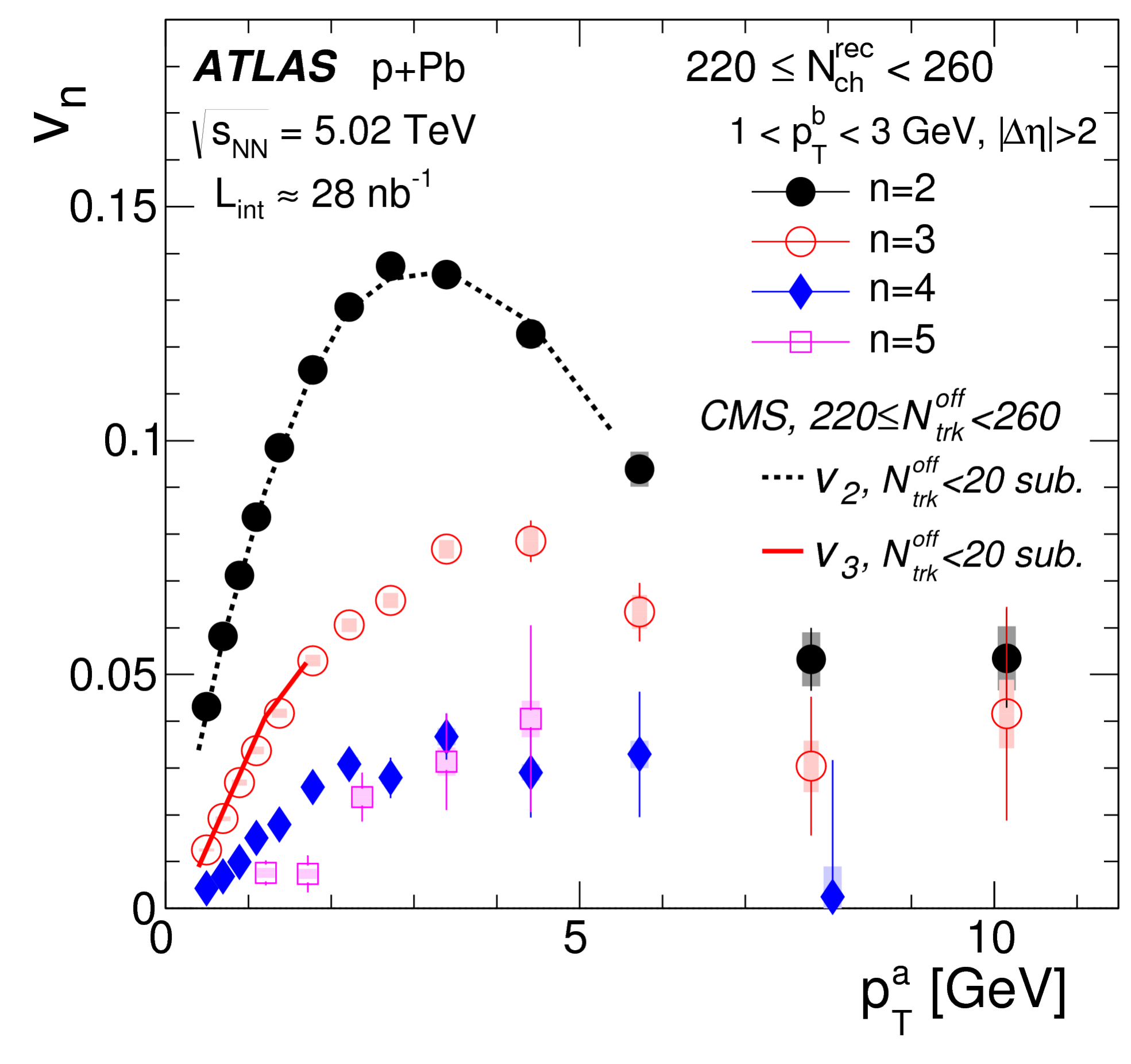} \label{fig:ATLAS_vn}
}
\subfigure[]{
\includegraphics[width=0.45\textwidth]{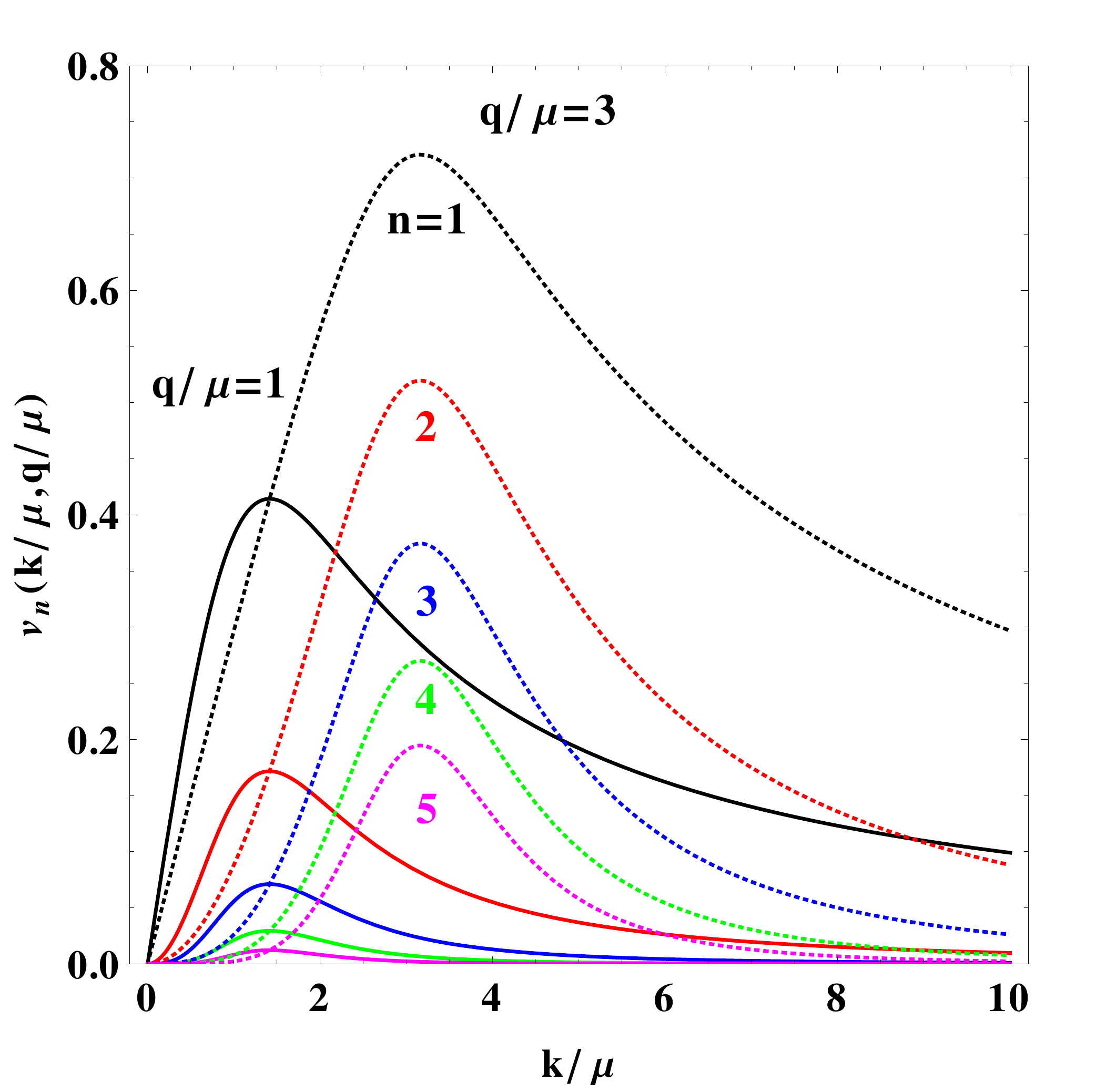} \label{fig:gluon_vn}
}
\caption{Left panel: Experimental $v_n$ harmonics for p+Pb collisions and $|\Delta \eta|>2$ taken from ATLAS~\cite{ATLASvn}. The magnitudes of the moments are comparable to the one observed in A+A reactions. Right panel: Azimuthal Fourier moments arising from the GB distribution (Eq.~\eqref{eq:GB}) as a function of the gluon transverse momentum and for different choices of the exchanged momentum, as computed in~\cite{Gyulassy}.} \label{fig:vn_intro}
\end{figure}

The recent discovery of long-range p+A azimuthal harmonics , see Fig.~\ref{fig:ATLAS_vn},
with magnitudes and $p_T$ dependence comparable to the ones found in A+A
collisions~\cite{CMS:2012qk, Abelev:2012ola, ATLASpA, Adare:2013piz},
 and the near energy independence of these
A+A moments observed in the Beam Energy Scan (BES) at
RHIC~\cite{BES} together with LHC have challenged the uniqueness of
the sQGP interpretation of $v_n$ in A+A. Prior to the recent p+A and BES data,
it was assumed that in smaller transverse size p+A systems or lower
energy A+A reactions, the perfect fluid sQGP ``core'' would gradually
change into a highly dissipative hadron resonance gas ``corona'' 
and lead to a substantially smaller magnitude of the azimuthal harmonic moments.
The observed near independence of  $v_n(p_T)$ to system size and initial energy density has motivated the search for possible alternative non-hydrodynamic 
sources of azimuthal harmonics.

While it has been shown that hydrodynamic equations, with particular
assumptions about the initial and freeze-out conditions~\cite{Bozek},
are {\em sufficient} to describe the data, the {\em uniqueness} of that
description is not obvious, especially given the unexpected features of the previously mentioned data.
One important class of non-hydrodynamic models proposed 
to explain this puzzle 
is based on 
the Color Glass Condensate (CGC) and Glasma paradigm
involving initial-state non-perturbative classical field correlations
controlled by a gluon saturation scale, $Q_\text{sat}(x,A)$
\cite{Raju,Kovner:2012jm,McLerran:2014uka,Levin,Ozonder:2014sra,Kovchegov:2012nd,Dumitru,JalilianMarian:1996xn,Skokov}. A simpler perturbative 
QCD source of multi-gluon azimuthal correlations due to classical non-abelian field interference effect  was recently presented in GLVB~\cite{Gyulassy} based on the well-known Gunion-Bertsch (GB) LO analytic formula~\cite{Gunion:1981qs}. This non-abelian gluon radiation generates non-zero $v_n$ moments with a shape that closely resembles the experimental one --- see Fig.~\ref{fig:gluon_vn}. The magnitudes of those parton level harmonics are however too large and some kind of damping mechanism is therefore required. A natural candidate for this job is hadronization. In fact, the primary advantage of GLVB, for our purpose
of exploring systematic theoretical uncertainties associated with
hadronization scheme dependence of azimuthal harmonics, is
 its ease of adaptability to Monte Carlo (MC) multi-particle production in p+p, p+A and A+A collisions via the HIJING~\cite{HIJING} type of models as emphasized in~\cite{Gyulassy}.

Thus far  CGC/Glasma phenomenology of azimuthal harmonics moments in p+A and A+A reactions has been limited to the idealized local parton-hadron hadronization scheme~\cite{Azimov:1984np} that by assumption preserves the parton level correlations into the final-state 
hadrons.  This guarantees minimal distortion of the initial-state multi-parton correlations predicted by QCD. More realistically, it is well-known from decades of hadronic phenomenology that phenomena like the production and dacay of intermediate hadron resonances and
final-state correlations can introduce non-trivial complications and, therefore, some non-perturbative hadronization scheme is required.
Hadronization phenomenology cannot be rigorously predicted from QCD, 
but two generic approaches, carefully tuned to $e^+e^-$, $ep$, and $pp$ data, have been developed during the years. These models are the independent fragmentation scheme~\cite{Field:1977fa} and the LUND string model~\cite{Andersson:1986gw} implemented in the JETSET algorithm~\cite{Sjostrand:1993yb}. They can be used to estimate the distortions of initial-state partonic correlations due to a more realistic hadronization process.
In particular, in the nuclear event generator HIJING, JETSET is used to convert the multiple diquark-quark beam jets with gluon mini-jet kinks into the multi-hadron resonances with particle data book properties
and decay branchings.

In this paper we study the hadronization scheme dependence of final-state
correlations using three different hadronization models that can be conveniently selected within the JETSET code. We deliberately neglect any intermediate parton or hadron level transport or hydrodynamic
effect to concentrate exclusively on final-state
hadronization modification of initial multi-gluon correlations.
We compare: $(1)$ local parton-hadron
duality (LPH), $(2)$ collinear (independent) parton-to-hadron resonance fragmentation/decay (CPR) and
$(3)$ LUND string hadronization models. We treat
$Q_\text{sat}$ as an input parameter controlling the $p_T$ range of azimuthal
asymmetry of Gunion-Bertsch multi-beam gluon bremsstrahlung. Our
results strongly support the analysis of Skokov \emph{et al.}~\cite{Skokov},
indicating that the CGC/Glasma predictions (without detailed
hadronization modelling) should be limited to the kinematic range $p_T>3$ GeV
to avoid complications and theoretical uncertainty of non-perturbative hadronization physics.

At this point, it should be stressed that the configurations that we will consider in this work are quite simple and hence cannot be considered as a realistic representation of the ridge effect observed experimentally in p+A reactions. However, as we will see, they are already enough to draw strong conclusions about the hadronization mechanism dependence of the final azimuthal harmonics.

\section{Elliptic interference harmonics of gluons from two recoiling beam jets}
\label{sec:simulation}

For our simulations, we used the MC hadronization algorithm JETSET 7.4~\cite{Sjostrand:1993yb} with $30\times10^6$ simulated GLVB  ``events'' with two
recoiling beam jets. Each beam jet is represented by 
a high invariant mass $q\bar q$ pair along a ``beam axis'', $\hat{z}$.  
For our simulations the invariant mass of each beam jet 
 is taken to be $100$ GeV. 
The two  beam jets are assumed to scatter with equal but opposite
net momentum transfer $\vett{Q}_1=-\vett{Q}_2=(q,\psi)$, with magnitude
distributed according to a simple Rutherford form $Q_{sat}^2/(q^2+Q_{sat}^2)^2$, and azimuthal direction, $\psi$, distributed uniformly in $[-\pi,\pi]$.
For our numerical simulations,  we further 
assumed Poisson fluctuations of the number, $N_g$, of bremsstrahlung gluons
with $\langle N_g\rangle=8$ and distributed uniformly in rapidity between 
kinematic bounds. 

In the LUND color string model~\cite{Andersson:1986gw}, 
all gluons are represented by ``kinks'' of the $q\bar{q}$ string
that deform it in the transverse plane. In contrast, 
from an independent fragmentation point of view~\cite{Field:1977fa},
gluons  are assumed to be just isolated partons 
that hadronize independently from
each other. We generate the bremsstrahlung gluons 
transverse momenta and azimuthal angles, $\vett{k}=(k,\phi)$,
from the perturbative regularized GB distribution:
\begin{align} \label{eq:GB}
f_{GB}^\text{reg}(\vett{Q},\vett{k})\propto\frac{Q_\text{sat}^2}{\left(q^2+Q_\text{sat}^2\right)^2}\frac{q^2+\mu^2}{\left(k^2+\Lambda_{QCD}^2\right)\left(k^2+q^2+\mu^2-2kq\cos(\phi-\psi)\right)},
\end{align}
with $Q_\text{sat}=2$ GeV being the typical momentum scale expected
from the CGC model for p+A collisions. We take $\mu=300$ MeV and $\Lambda_{QCD}=200$ MeV as infrared regulating scales.

 This setup produces initial azimuthal anisotropy at the parton level 
that with the simplest LPH scheme leads to harmonic moments
similar to experiment, as seen in Fig.\ref{fig:vn_intro}.
In particular, the pseudo-rapidity independence and the preference of radiated gluons transverse momenta $\vett{k}$ 
to be aligned
along $\pm \vett{Q}$ gives rise to a  ``ridge-like'' structure with azimuthal asymmetry 
resembling the one observed experimentally.

\begin{figure}[t]
\centering
\includegraphics[width=\textwidth]{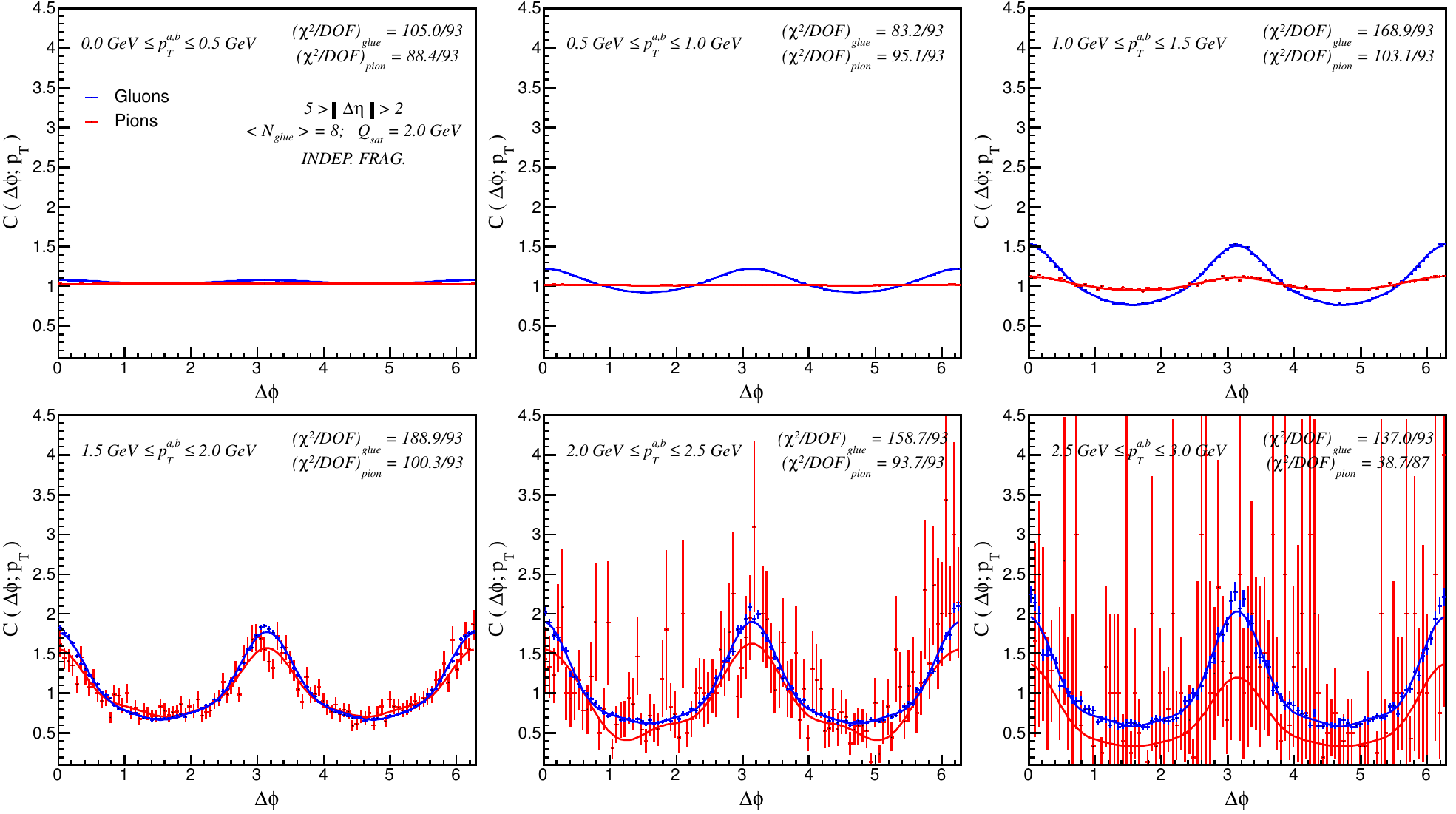}
\caption{$\Delta \phi$ distributions for pairs of initial gluons (blue) and final pions (red) for different $p_T$ bins. The hadronization has been performed using the independent fragmentation scheme. The solid lines are the fitted curves.} \label{fig:distrInd}
\end{figure}

\begin{figure}[t]
\centering
\includegraphics[width=\textwidth]{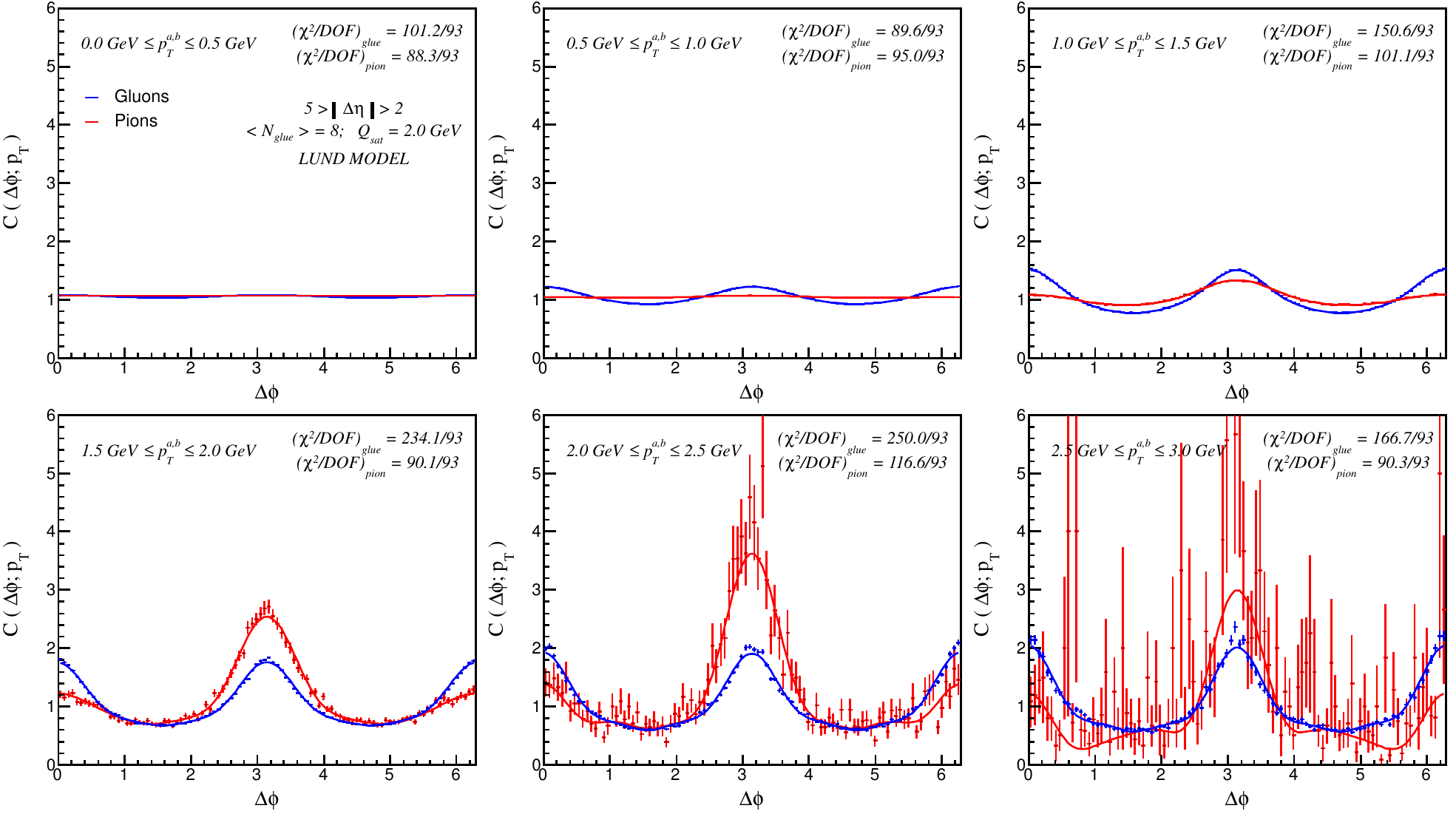}
\caption{Same quantities as reported in Fig.~\ref{fig:distrInd} but with hadronization performed using the LUND string model.} \label{fig:distrLund}
\end{figure}

After the hadronization, an analysis is performed over pairs of final 
pions with criteria as close as possible to~\cite{ATLASpA}. 
In particular, our pseudo-rapidity cuts on the final pions are $|\eta|<2.5$ and $|\Delta\eta|>2$ (the so-called long-range), $\Delta\eta=\eta_a-\eta_b$ being the relative psuedo-rapidity of the pair. Both pions are taken in the same $p_T$ bin and the correlation functions we computed are given by:
\begin{align}
C(\Delta \phi;p_T)=\frac{S(\Delta \phi;p_T)}{B(\Delta \phi;p_T)},
\end{align}
where $\Delta\phi=\phi_a-\phi_b$ is the relative azimuthal angle between the two pions and $S$ and $B$ represent the same and mixed event pair distributions respectively~\cite{ATLASpA}. The analysis is repeated for the three
hadronization schemes to reveal the hadronization scheme 
dependence of final azimuthal harmonics shown in Figs.~\ref{fig:distrInd} and \ref{fig:distrLund}.

\begin{figure}[t]
\centering
\subfigure{
\includegraphics[width=0.47\textwidth]{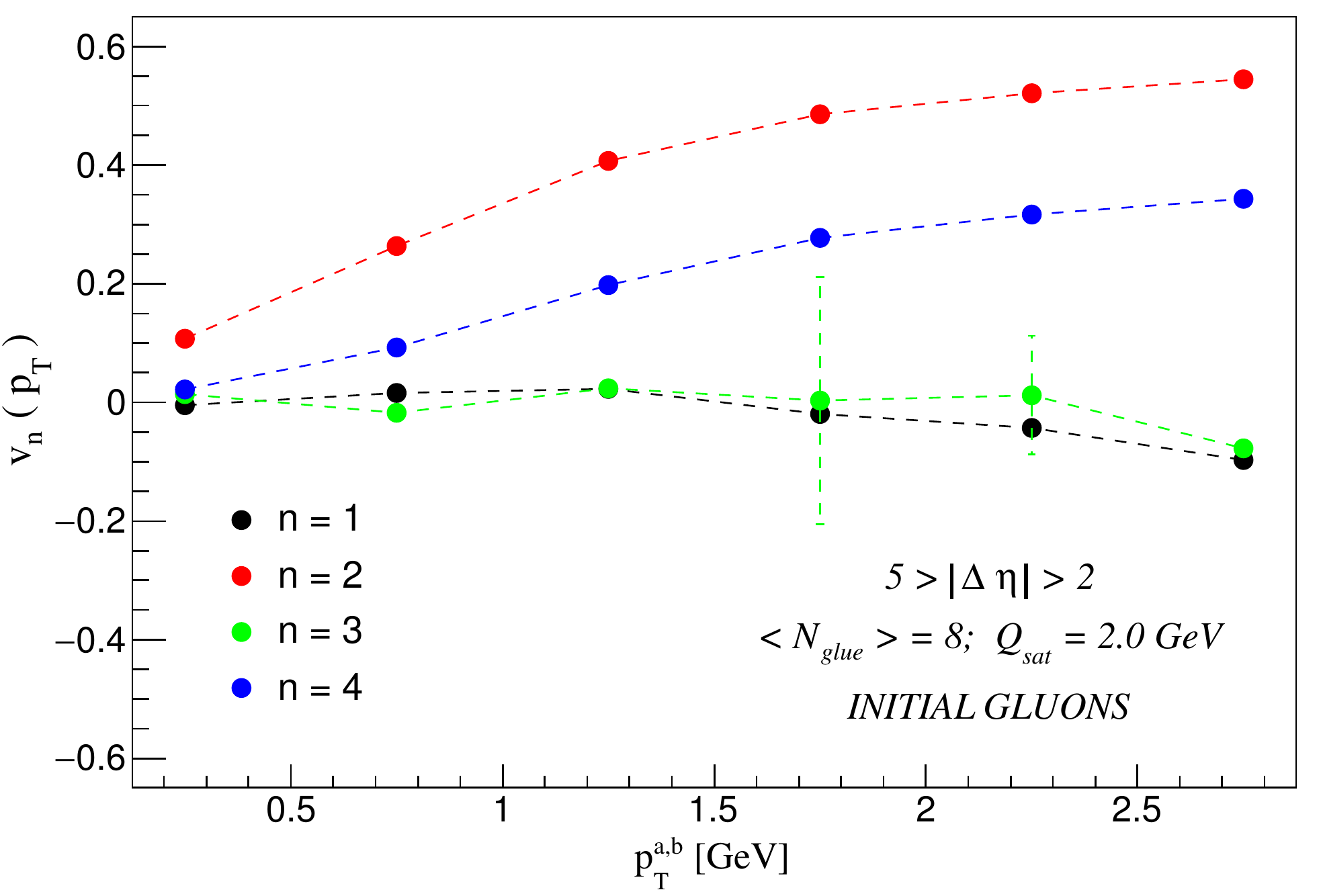} \label{fig:vnGlue}
}

\subfigure{
\includegraphics[width=0.47\textwidth]{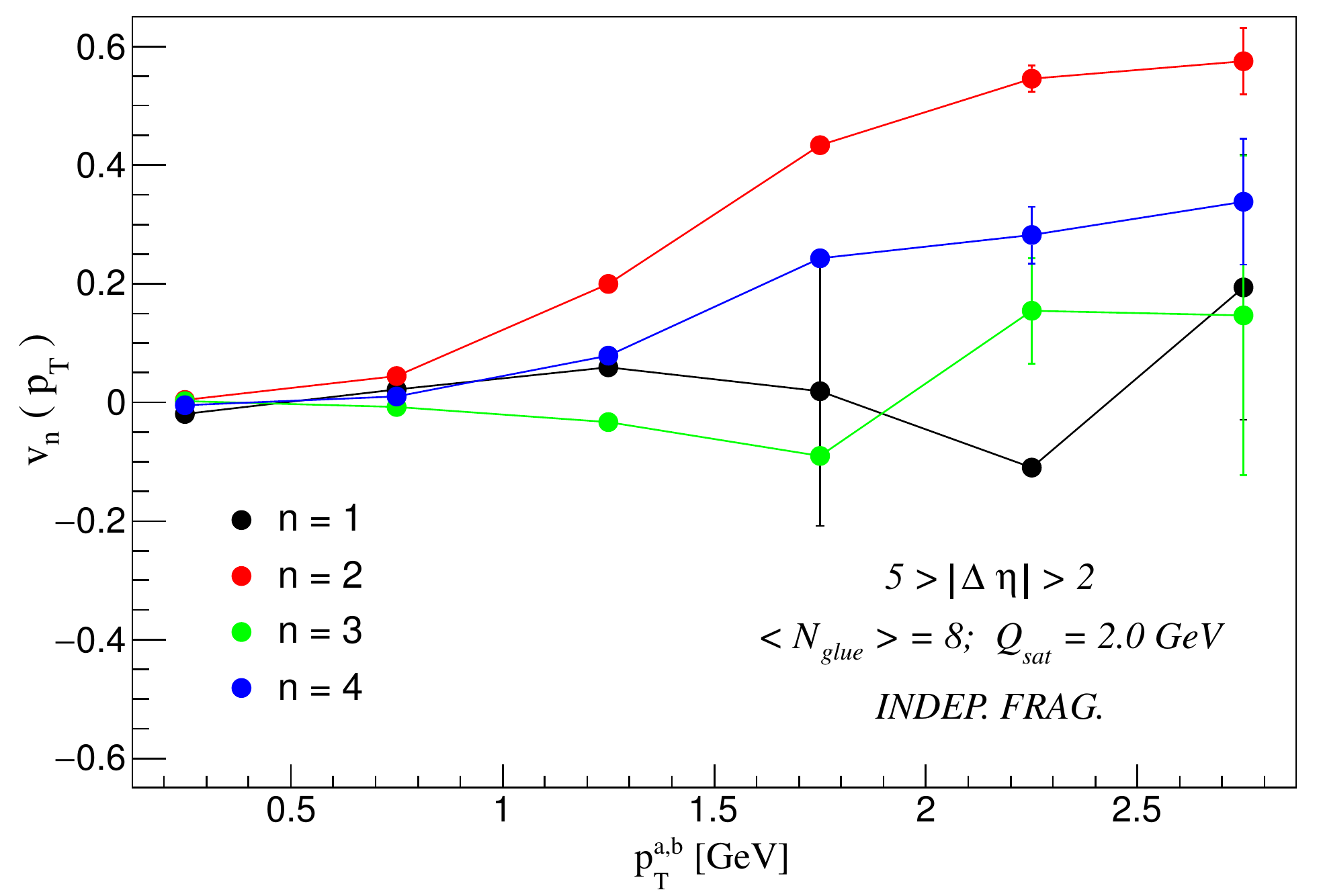} \label{fig:vnInd}
}
\subfigure{
\includegraphics[width=0.47\textwidth]{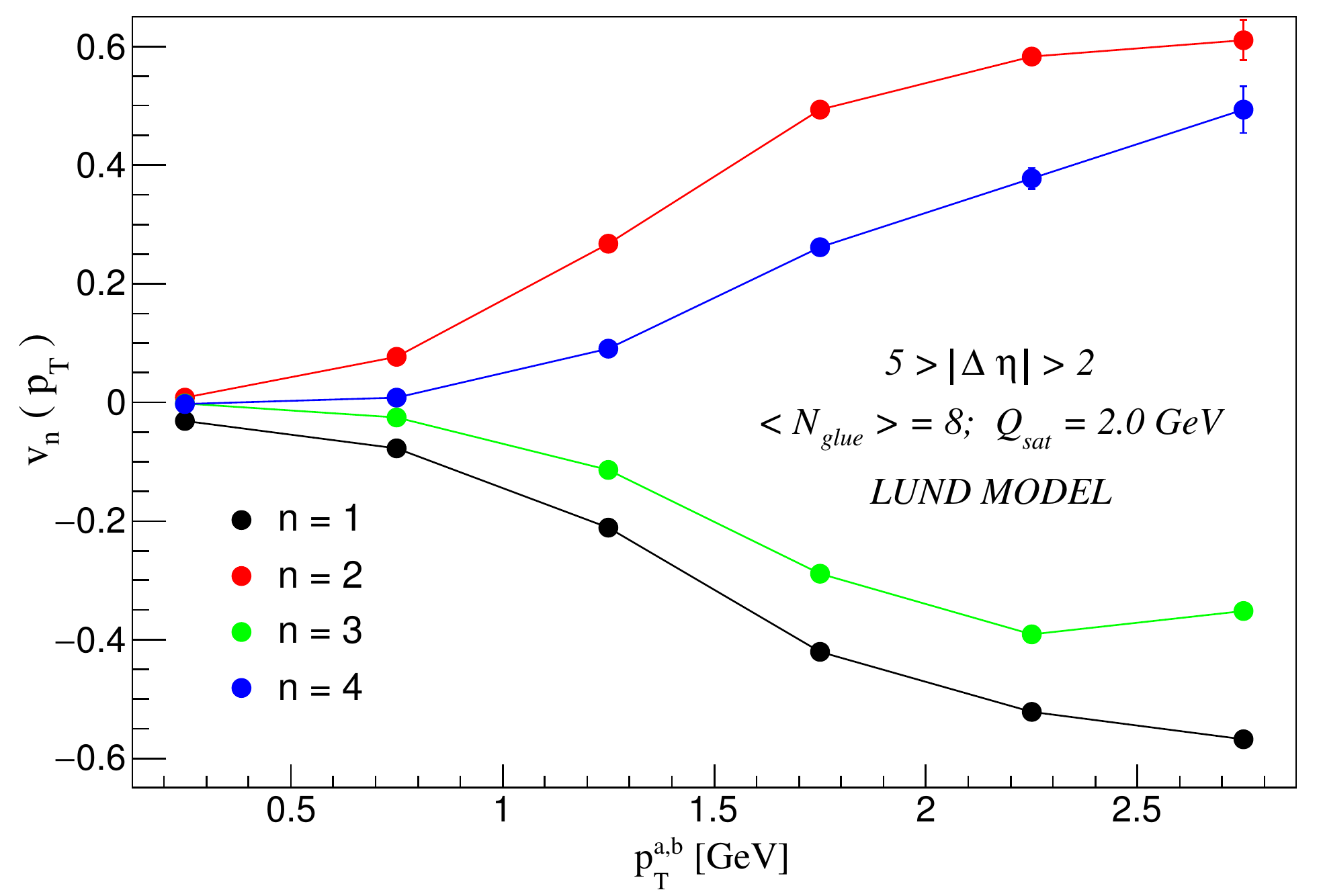} \label{fig:vnLund}
}
\caption{Fitted single-particle azimuthal moments, $v_n(p_T)$, for initial gluons (upper panel) and final pions as obtained from the independent fragmentation scheme (lower left panel) and from the LUND model (lower right panel). The gluons do not show any odd Fourier coefficient. This feature is shared by the independent fragmentation case as well while the Lund model case presents large negative $v_1$ and $v_3$.} \label{fig:vn}
\end{figure}

From these distributions we extracted the $n\leq6$ single-particle moments, $v_n(p_T)$, using the following fitting function:
\begin{align}
C_\text{fit}(\Delta\phi;p_T)=C_0\left(1+2\sum_{n=1}^6v_n^2(p_T)\cos(n\Delta\phi)\right).
\end{align}
The results of this fit for $n\leq4$ are shown in Fig.~\ref{fig:vn}.

\subsection{Numerical Results}
\label{sec:results}

The computation of the two-particle $\Delta\phi$ distributions as reported in Figs.~\ref{fig:distrInd} and \ref{fig:distrLund} shows that the process of hadronization has some deeply non-trivial consequences for the correlations between final state particles.

The gluon curves clearly present the behavior expected from the discussion around Eq.~\eqref{eq:GB}. The long-range near-side ($\Delta\phi\simeq0$) peak is simply due to the fact that the gluons are produced with transverse momentum preferentially close to the exchanged momentum $\vett{Q}$, while the away-side ($\Delta\phi\simeq\pi$) peak is produced by the recoil of the two strings. These features extend over the full $p_T$ range.

The distributions of the final pairs of hadrons, instead, have some very peculiar properties, depending on which $p_T$ region we are considering. For small values of the pion transverse momentum --- smaller than a scale, $\lambda$, to be determined in Sec.~\ref{sec:scale} --- the initial anisotropy is extremely reduced and the two-pion distributions become more and more uniform for decreasing $p_T$. This behavior is common to both the LUND model and the independent fragmentation model, even though the shape of the $\Delta\phi$ distributions already appears to be different for $p_T>1$ GeV. The reasons for this strong dilution of the initial anisotropy are proper to the process of hadronization itself and will be analyzed again in Sec.~\ref{sec:scale}. 
If a simple system like ours already has such a ``decoherence power'', one can reasonably expect the same thing to be valid for a more realistic situation --- see for example Sec.~\ref{sec:Mercedes}. It should also be stressed that, since the low-$p_T$ pions seem to carry no information about the initial gluon distribution, this feature appears be true in general, no matter what causes the partonic correlation in the first place, and hence can be applied to other initial-state models as well.

The second result refers to the higher transverse momentum regime ($p_T\gtrsim \lambda$). In both models this region presents strong two-particle correlations among the final pions, since the gluon anisotropy is transmitted more efficiently. However, one can immediately appreciate an essential difference between the independent fragmentation and the QCD string models. In the former, by definition, each parton fragments independently from the others and hence the pion correlation function closely resembles the gluon one since, when higher $p_T$ are considered, no other relevant sources of angular correlation come into play. For the case of the LUND model, instead, while the near-side peak is generally lower for the pions than for the gluons, an extremely pronounced away-side peak is present. This large $\Delta\phi\simeq\pi$ signal (several times bigger than the $\Delta\phi\simeq0$ one) is a direct consequence of the transverse momentum conservation taking place at each breaking of the color string~\cite{Sjostrand:1993yb}. Each $q\bar q$ system, in fact, has zero total $p_T$ and the transverse fluctuations are governed by a roughly Gaussian probability distribution. This means that when the string breaks and a new pair of partons, say $q^\prime$ and $\bar q^{\,\prime}$, is created from the vacuum, the $p_T$ of these partons has a probability distribution given by
\begin{align} \label{eq:exp}
\mathcal{P}(p_T)\propto \exp\left(-\frac{\pi m_T^2}{\kappa}\right)=\exp\left(-\frac{\pi m^2}{\kappa}\right)\exp\left(-\frac{\pi p_T^2}{\kappa}\right),
\end{align}
with $m$ being the mass of the produced $q^\prime$ and $\kappa\simeq1$ GeV/fm being the string constant. By conservation of total momentum they are always produced back-to-back. It then follows that, for every hadron of momentum $\vett{p}_T$ there will always be another with momentum $-\vett{p}_T$, \emph{i.e.} such that $\Delta\phi=\pi$. This essentially means that the differential distribution $C(\Delta\phi;p_T)$ contains a contribution proportional to $\delta(\Delta\phi-\pi)$, which gets broader when finite $p_T$ bins are taken into account, as in any realistic situation. We conclude that the QCD strings hadronization scheme 
that conserves event
energy-momentum (unlike independent fragmentation) 
can introduce large away-side correlations in the spectrum of the final hadrons
that can lead to negative odd harmonics absent at the 
purely partonic level.

All these properties lead to the azimuthal moments, $v_n(p_T)$, shown in Fig.~\ref{fig:vn} which therefore differ dramatically between the two hadronization schemes. In particular, as a consequence of the almost complete flatness of the $\Delta\phi$ distributions, both models present even harmonics for low-$p_T$ which are much smaller than the gluonic ones.  To be more quantitative, in Tab.~\ref{tab:vnratio} we compare the magnitudes of the even azimuthal harmonics at the partonic and hadronic level. As one can see, the damping effect is almost total for $p_T<1$ GeV and persists up to $p_T\lesssim 2$ GeV.

 Moreover, the prominent away-side peak present in the distributions of Fig.~\ref{fig:distrLund} causes large negative odd harmonics due to assumed local transverse momentum conservation. 
 
It should also be noted that these $v_n$ arise solely from  initial GB gluon bremsstrahlung and final hadronization effects. No final state interactions have been included in our simulations. 

\begin{table}[t]
\centering
\begin{tabular}{c|cccc||cccc}
\hline\hline
& \multicolumn{4}{c||}{Independent fragmentation} & \multicolumn{4}{c}{LUND model} \\
\hline
$p_T$ [GeV] & 0.25 & 0.75 & 1.25 & 1.75 & 0.25 & 0.75 & 1.25 & 1.75 \\
\hline\hline
$v_2^{\,\pi}/v_2^{\,g}$ & 0.04 & 0.10 & 0.36 & 0.60 & 0.06 & 0.30 & 0.65 & 1.00 \\
\hline
$v_4^{\,\pi}/v_4^{\,g}$ & 0.00 & 0.03 & 0.25 & 0.63 & 0.00 & 0.20 & 0.50 & 1.00 \\
\hline\hline
\end{tabular}
\caption{Ratios between the pion even moments ($v_n^{\,\pi}$) and the gluon ones ($v_n^{\,g}$) for $p_T<2$ GeV. One immediately notices that there is a consistent damping of the initial magnitudes. In particular, the final-state $v_n$ are essentially zero (less than $30\%$ of the initial ones) for $p_T<1$ GeV. The suppression for $p_T>2$ GeV appears to be negligible.} \label{tab:vnratio}
\end{table}

\section{Hadronization  mechanisms that damp gluon harmonics} 
\label{sec:scale}

At this point one needs to ask: what is the scale which determines the quenching of anisotropy for small values of the transverse momenta? Our picture has two natural scales: the typical energy of the hadronization process, $\lambda\simeq1$ GeV, and the typical momentum exchanged, $Q_\text{sat}=2$ GeV. Which of the two plays the role of discriminant between the uniform and the anisotropic regions?

\begin{figure}[t]
\centering
\includegraphics[width=\textwidth]{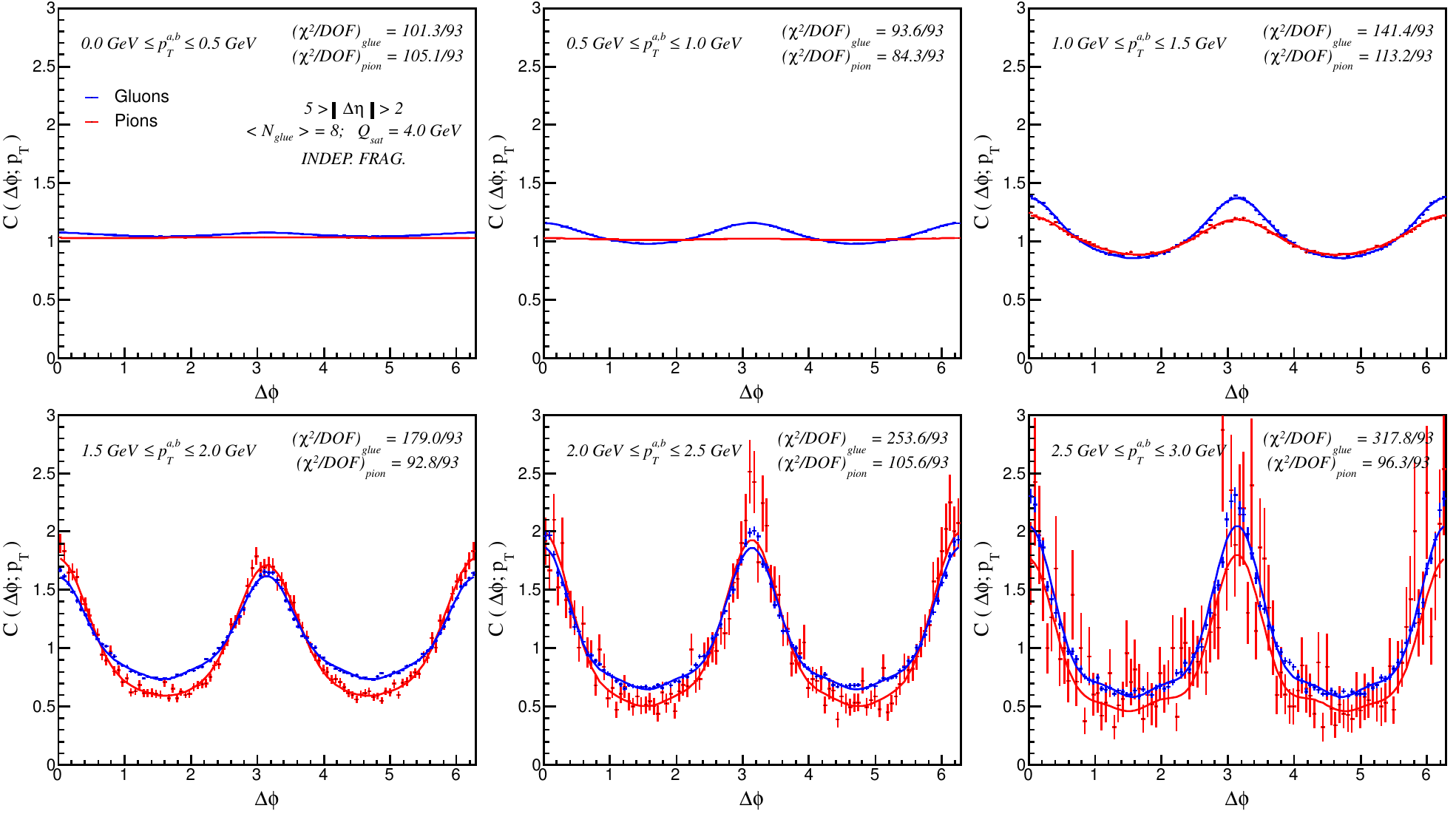}
\caption{Same quantities as in Fig.~\ref{fig:distrInd} and \ref{fig:distrLund} but obtained using the independent fragmentation scheme and $Q_\text{sat}=4$ GeV. It is evident how increasing the momentum exchanged enhances both the initial and final anisotropies for every value of the trasnverse momentum except for $p_T\leq \lambda =1$ GeV (first two upper panel from the left), where the pions distributions are completely uniform.} \label{fig:Qsat4}
\end{figure}

To answer this question we repeated the previous simulation using the independent fragmentation scheme but with a different value for the average momentum exchanged, $Q_\text{sat}=4$ GeV. The results are reported in Fig.~\ref{fig:Qsat4}.
If compared to Fig.~\ref{fig:distrInd} one can appreciate that the final hadronic anisotropies now follow the gluonic ones much more closely than in the $Q_\text{sat}=2$ GeV case, as one might expect, for almost all values of the transverse momentum except for $p_T\leq 1$ GeV, where the two-pion distributions are again completely flattened out. This shows that the energy scale relevant in this decoherence effect is indeed the scale of hadronization, $\lambda \simeq 1$ GeV.

As a further check, one can study the average ratio between the pion and the gluon momenta both for the transverse and longitudinal components. Our simulation indicates that for the two models
\begin{align}
\frac{\left\langle p_T^{\,\pi} \right\rangle}{\left\langle p_T^{\,g}\right\rangle}=\begin{cases}
0.54 & \text{Indep. fragm.} \\
0.58 & \text{LUND}
\end{cases}
;\quad\quad \frac{\left\langle p_z^{\,\pi} \right\rangle}{\left\langle p_z^{\,g}\right\rangle}=\begin{cases}
2.08 & \text{Indep. fragm.} \\
2.51 & \text{LUND}
\end{cases}.
\end{align}
This again is a hint for the role of hadronization in damping the anisotropies at low $p_T$.

The reasons for this ``quenching power'' of the hadronization process can be found in essentially two features: \emph{(i)} non-collinearity of the gluon radiation and \emph{(ii)} isotropic decays of resonances. In particular, the fact that gluon fragmentation is not perfectly collinear is already manifest in the first order pQCD Altarelli-Parisi distribution,
\begin{align}
\frac{dN_g}{dz d\vett{k}^2}=\frac{\alpha}{2\pi}\frac{1}{\vett{k}^2}P(z),
\end{align}
where $\vett{k}$ is again the gluon transverse momentum, $z$ is the Bjorken variable and $P(z)$ the splitting function. Although this distribution is strongly peaked around $\vett{k}\simeq0$ it also has very long tails, clearly showing that a totally collinear picture is a too na\"{i}ve approximation (for an interesting discussion on the role of fragmentation functions on the azimuthal harmonics see~\cite{Torrieri:2013aqa}). Morever, in the LUND string model we also have the transverse fluctuations of the flux tube, whose typical scale is $\sqrt{\kappa}\simeq 0.45$ GeV (see Eq.~\eqref{eq:exp}), which provide another source of dilution of the initial anisotropy. Lastly, the intermediate steps between the initial gluons and the final pions are populated by resonances. These particles decay without any preferential direction and hence strongly contribute to the flattening of the final $\Delta\phi$ distributions. In particular, this is the reason why the observed decoherence happens for $p_T\lesssim1$ GeV, this being the typical mass of the most common resonances. To better illustrate this point we performed an overly simplified simulation whose description and results are shown in Tab.~\ref{tab:1glue}.

\begin{table}[h!]
\centering
\begin{tabular}{c|ccc||c|ccc}
\hline\hline
\multicolumn{4}{c||}{Independent fragmentation} & \multicolumn{4}{c}{LUND model} \\
\hline
\multicolumn{8}{c}{Event 1}\\
\hline
Particle & $p_T$ [GeV] & $p_z$ [GeV] & $\phi$ [rad] & Particle & $p_T$ [GeV] & $p_z$ [GeV] & $\phi$ [rad] \\
\hline
$(\Delta ^{++})$ & 0.90 & 0.90 & 1.74 & $\pi^-$ & 0.22 & 0.17 & 2.67 \\
$(\,\rho^0)$ & 1.00 & 0.36 & 0.19 & $(\omega)$ & 1.20 & 0.00 & 0.09 \\
$(\,\rho^+)$ & 0.56 & -0.38 & -0.53 & $(\,\rho^0)$ & 1.11 & 0.91 & -0.02 \\
$p$ & 0.79 & 0.78 & 1.95 & $\pi^-$ & 0.77 & -0.03 & 0.01 \\
$\pi^+$ & 0.20 & 0.11 & 0.79 & $\pi^+$ & 0.39 & 0.11 & -0.74 \\
$\pi^+$ & 0.44 & -0.01 & -0.48 & $\pi^0$ & 0.11 & 0.19 & 0.76 \\
$\pi^-$ & 0.71 & 0.37 & 0.58 & $\pi^+$ & 0.15 & 0.19 & 0.76 \\
$\pi^+$ & 0.59 & -0.03 & -0.35 & $\pi^-$ & 0.29 & 0.49 & -0.09 \\
\hline
\multicolumn{8}{c}{Event 2} \\
\hline
$(\,\rho^0)$ & 0.26 & 0.08 & 0.67 & $(\,\rho^+)$ & 0.57 & 0.36 & 0.72 \\
$\eta$ & 2.15 & -0.22 & 0.04 & $(\,\rho^+)$ & 0.57 & 0.36 & 0.72 \\
$\pi^+$ & 0.49 & -0.39 & -1.34 & $(\,\rho^-)$ & 0.81 & 0.84 & 0.59 \\
$K^0$ & 0.42 & -0.24 & -1.55 & $(\,\rho^0)$ & 0.97 & -0.07 & -0.56 \\
$\pi^-$ & 0.39 & 0.10 & 3.06 & $\pi^+$ & 0.06 & 0.07 & -2.86 \\
$\pi^+$ & 0.33 & 0.23 & -0.31 & $\pi^-$ & 0.68 & 0.35 & -0.92 \\
$\pi^+$ & 0.47 &-0.16 & 0.54 & $\pi^+$ & 0.19 & 0.09 & 2.51 \\
$\pi^-$ & 0.21 & 0.24 & -2.77 & $\pi^0$ & 0.64 & 0.27 & 0.43 \\
$\pi^0$ & 0.26 & -0.11 & -2.96 & $\pi^-$ & 0.44 & -0.07 & 1.25 \\
$\pi^+$ & 0.41 & -0.29 & -3.04 & $\pi^+$ & 1.24 & 0.06 & -0.05 \\
\hline\hline
\end{tabular}
\caption{Output of two events composed by a $q\bar q$ pair and a gluon for both independent fragmentation and LUND model. The quark and antiquark are moving in opposite directions along $z$-axis, each of them with energy $E_q=10$ GeV. The gluon, instead, flies away along the $x$-axis ($\eta=0$, $\phi=0$) with energy $E_g=3$ GeV. We only report those particles with pseudo-rapidity $|\eta|<1$, \emph{i.e.} close to the gluon in phase space. One immediately notices that even though the initial gluon has zero azimuthal angle, the fragmentation produces particles with $\phi\neq 0$. Every event always contains at least one resonance (particles in parenthesis) which then decays isotropically. Moreover, the fact that even the resonances (which are produced in the very first step of hadronization) have non-zero anzimuthal angle shows that the gluon fragmentation is non-collinear. The final pions are therefore widely spread in $\Delta\phi$. They also have $p_T<1$ GeV as expected.} \label{tab:1glue}
\end{table}

We learn from above that the strong damping of parton level  $v_n$ at $p_T\lesssim\lambda$ is not a consequence of the particular initial-state model chosen, but it is rather an \emph{intrinsic} property of the hadronization mechanism. This is one of the striking results of our analysis.

\section{Hadronization damping of triangular and higher gluon harmonics}
\label{sec:Mercedes}
\begin{figure}[t]
\centering
\includegraphics[width=\textwidth]{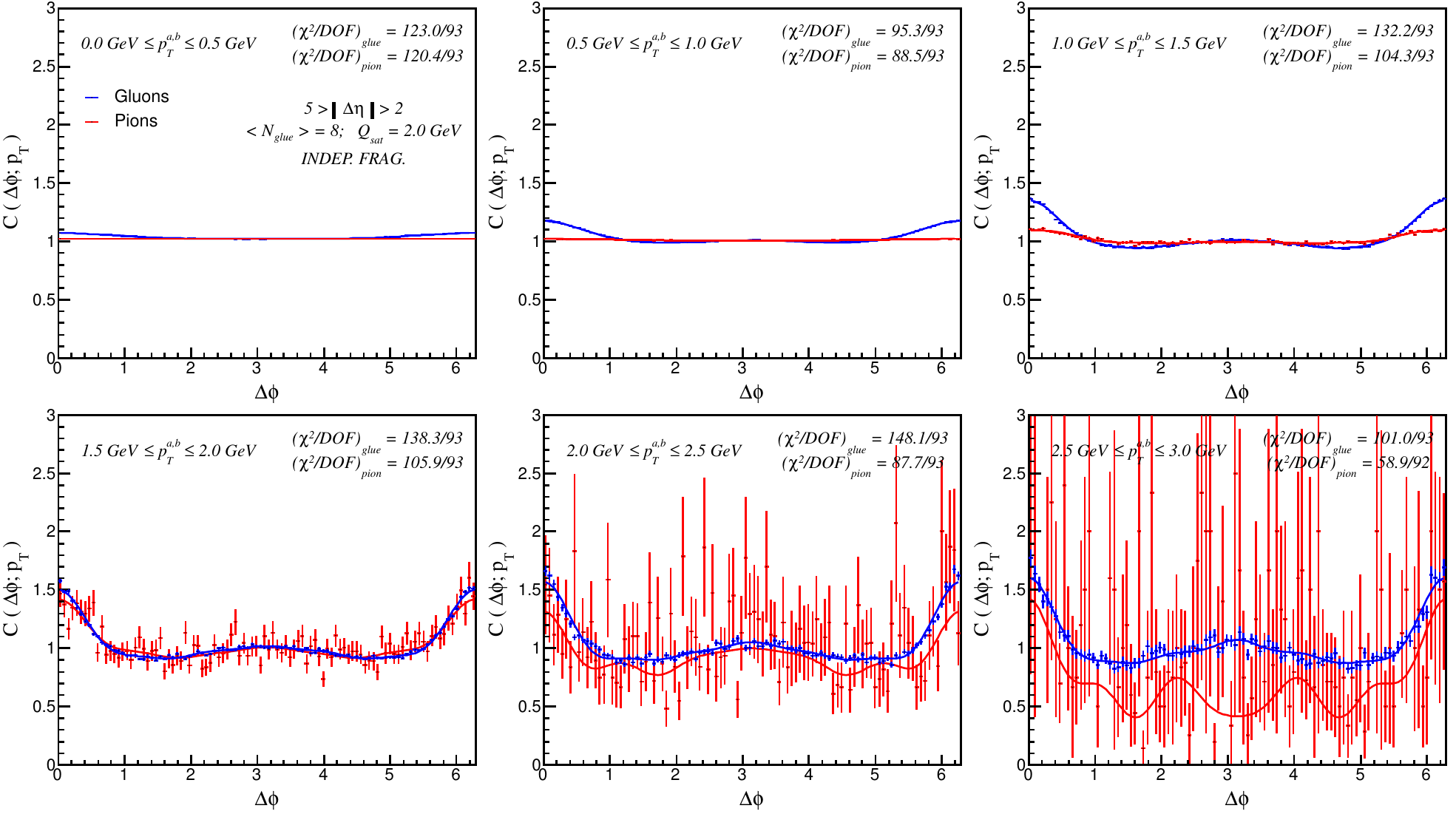}
\caption{Two-particle $\Delta\phi$ distributions obtained from three initial $q\bar q$ pairs in a triangular configuration for both the initial-state gluons (blue) and the final pions (red). The hadronization has been performed using the independent fragmentation scheme.} \label{fig:distrMercedes}
\end{figure}

\begin{figure}[t]
\centering
\subfigure{
\includegraphics[width=0.47\textwidth]{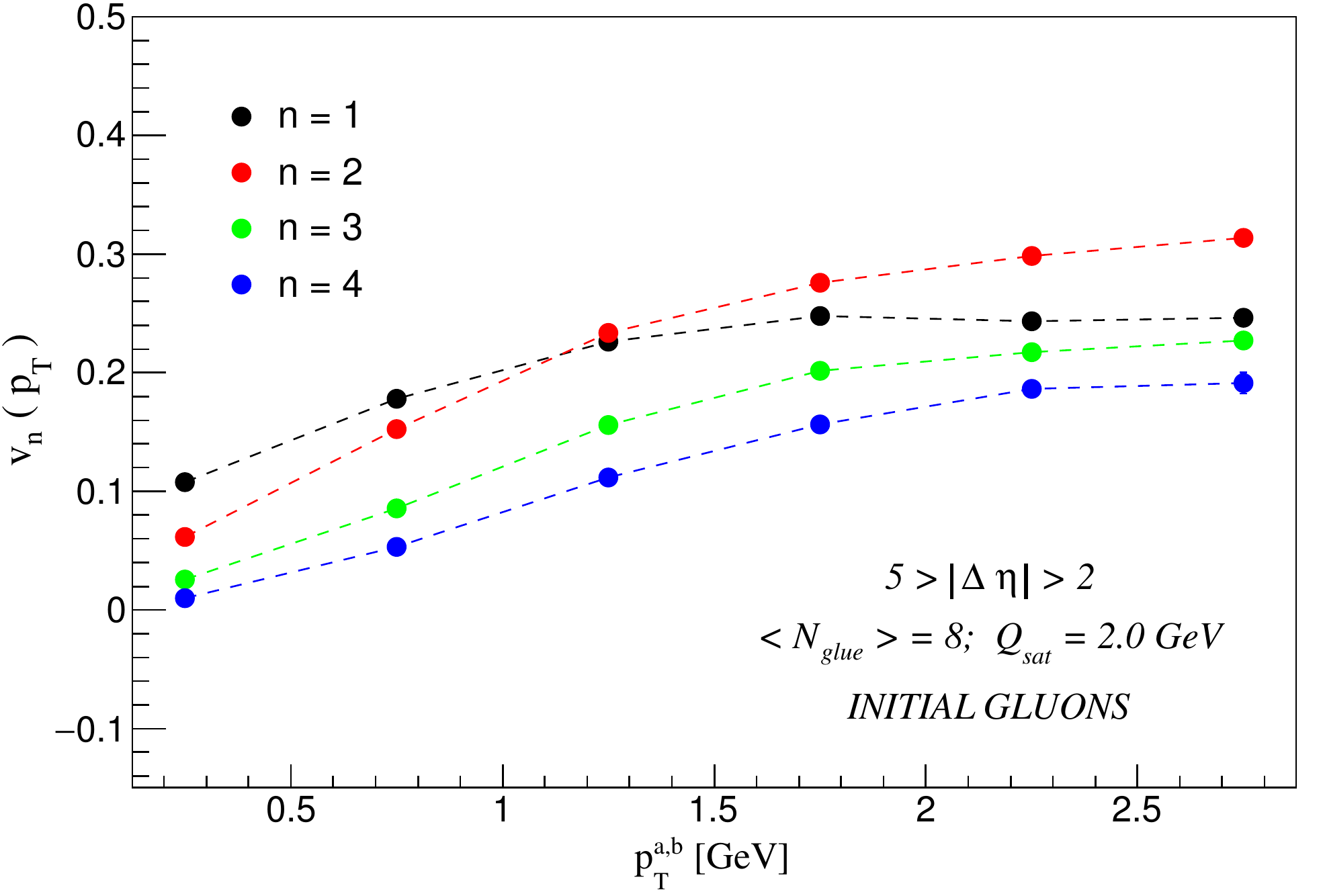}
}
\subfigure{
\includegraphics[width=0.47\textwidth]{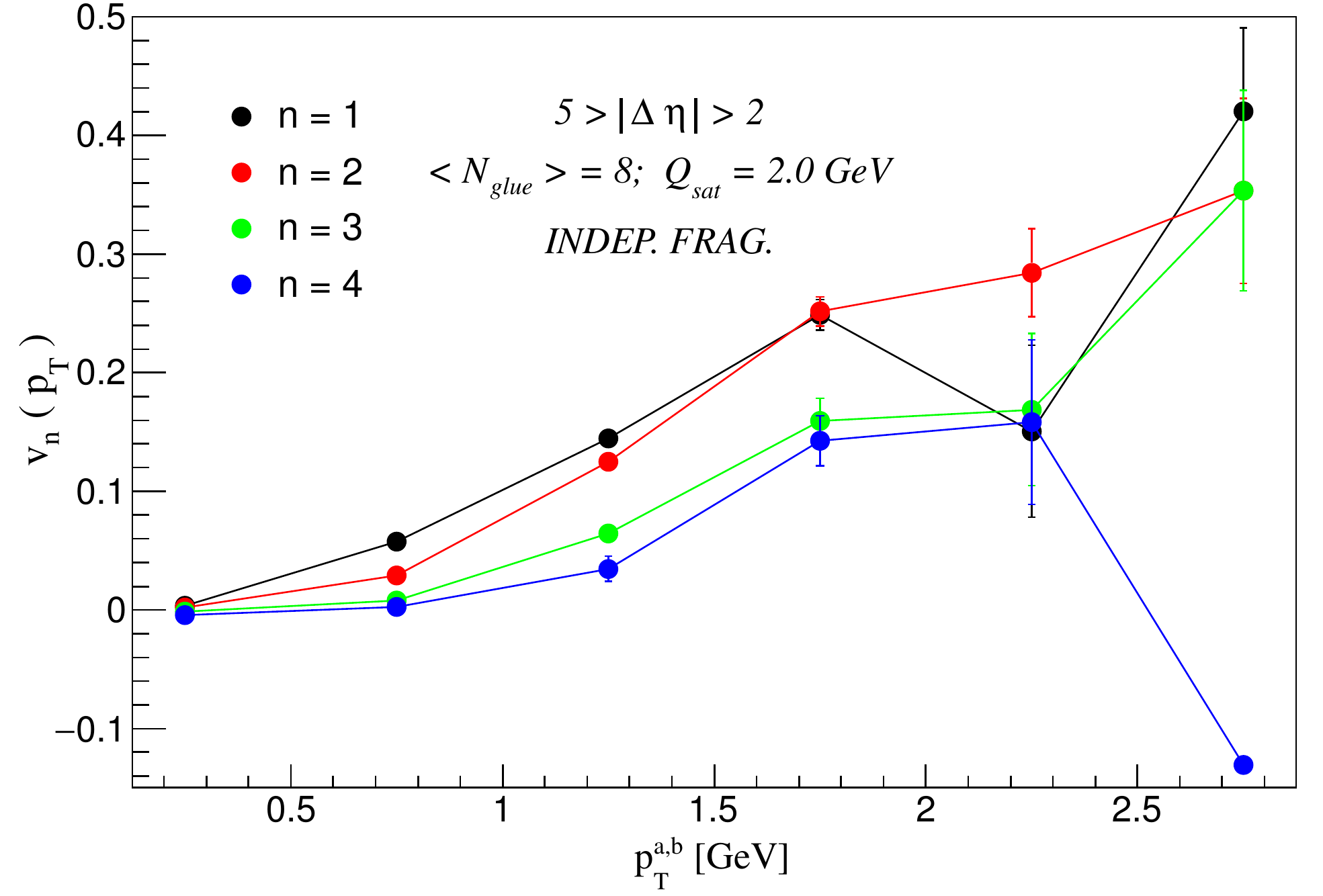}
}
\caption{Single-particle harmonic moments, $v_n(p_T)$, for initial-state gluons (left panel) and final-state pions (right panels) as extracted from the distributions in Fig.~\ref{fig:distrMercedes}. The higher-$p_T$ bins for the pions harmonics clearly suffer low statistics fluctuations.} \label{fig:vnMercedes}
\end{figure}

As one can see from Fig.~\ref{fig:vn}, in contrast to the LUND model, the independent fragmentation leads to very small odd harmonics --- \emph{albeit} fluctuations due to small statistics effects. The reason is that for two color antennas
the system is back-to-back symmetric, with just two $q\bar q$ pairs recoiling from each other. Since at high-$p_T$ no other correlations are introduced, the final pions inherit the symmetries of the initial partons. To check the consequences of hadronization on the purely geometrical odd moments --- and to reproduce a slightly more realistic configuration --- we performed a simulation involving three color antenna simulating three recoiling beam jets conserving transverse momentum.
 Both the simulation and the analysis closely follow what we explained in Sec.~\ref{sec:simulation} but now we implemented a third quark-antiquark pair such that the whole system conserves the total momentum, \emph{i.e.} if $\vett{Q}_1$, $\vett{Q}_2$ and $\vett{Q}_3$ are the momenta exchanged by each of the three pairs then $\vett{Q}_1+\vett{Q}_2+\vett{Q}_3=0$. In this case we only used the independent fragmentation scheme, which lacked the odd $v_n$ in the first place. The results for the $\Delta\phi$ distributions and for the final Fourier moments are shown in Fig.~\ref{fig:distrMercedes} and \ref{fig:vnMercedes}.

\subsection{Results}
As one immediately notices, the initial gluons $\Delta\phi$ distributions now have a contribution from odd Fourier components, $v_{2n+1}$. Once again, in the $p_T\lesssim \lambda$ region none of the initial information is preserved and the pions correlation functions are totally flattened. Even though we suffer from a lack of statistics, it is also clear that in the higher transverse momentum regime the final particles distributions keep following the gluonic ones. In terms of Fourier harmonics, even though the initial gluons have definitely sizeable $v_n$, the final pions moments in the low-$p_T$ region are essentially zero (less than $10\%$) --- see again Fig.~\ref{fig:vnMercedes}. Hadronization is once again very effective in quenching the azimuthal asymmetry of the system. For $p_T\gtrsim \lambda$ the gluonic and hadronic $v_n$ have, instead, similar values.

\section{Conclusions}
\label{sec:conclusion}
Our analysis shows that the process of hadronization can lead to major
distortions of the azimuthal harmonics up to $p_T=3$ GeV, which are of interest in the search for signatures of perfect fluidity in p+A systems. In particular, one important aspect seems to be general and
model-independent: at small values of the transverse momentum of the
hadrons the complexity of the hadronization itself --- namely
non-collinearity of fragmentation and isotropic resonance decays ---
greatly reduces the information contained in the initial-state
partons, causing an almost total smearing of the final hadronic
spectrum. As shown in Sec.~\ref{sec:scale}, this is intrinsic to the
hadronization itself and hence should be taken into account by every
model with initial-state anisotropy at least with a theoretical error
band estimated by testing several hadronization schemes.

Many models of initial-state anisotropy seem to claim that the only scale relevant for the generation of non-zero $v_n$ is the typical momentum exchanged, $Q_\text{sat}$. Our analysis shows that this is only true above
 a second hadronization scale, $\lambda\simeq1$ GeV.
Hadronization scheme choice \emph{does matter}, and those 
schemes that assume local collinear parton-hadron duality or pure collinear
fragmentation without resonance production may over-estimate final-state hadron harmonics. It is interesting to note that assuming that fragmentation is perfectly collinear and described by any fragmentation function $f(z)$ should actually lead to an enhancement of the initial gluon anisotropy 
since low-$p_T$ pions  come from gluons with higher transverse momentum 
that naturally have greater azimuthal asymmetry. This is in striking contrast with the results of our JETSET simulations,
where collinearity of hadronization is broken.

A second conclusion is that the two-particle correlations in the
$p_T\gtrsim \lambda \simeq 1$ GeV region can strongly depend on the
chosen hadronization model. Specifically, while in the independent
fragmentation scheme the initial parton anisotropy almost
completely transmits anisotropy to the final hadrons, in the LUND string model
a new source of correlation due to transverse momentum
conservation is introduced, leading to a large
away-side peak in the pion $\Delta\phi$ distributions that is not due to back-to-back mini jets, which are not taken into account in the present simulations. 
\emph{If} these
features also survive to a system more complicated than ours, then
this leads to two important remarks: first of all, whenever performing
a Monte Carlo simulation to study multi-particle correlations one
should always pay particular attention to the degree of
model-dependence of the simulation itself since this might introduce
important biases on the final conclusions. Secondly, if one assumes
the description of hadronization in terms of a QCD color string as a
fairly accurate one, then we might have an unexpected source of
\emph{non-flow} away side correlation for particles with $p_T>1$ GeV which
would be difficult to deconvolute from pQCD di-jets by any experimental analysis.

We emphasize again that the properties of the final two-hadron correlations computed in this study, for the most part, are due neither to a particular initial-state mechanism nor to any  collective transport or hydrodynamic effects: they are genuine consequences of the hadronization process that are scheme dependent.
There is unfortunately  no guarantee of universality of hadronization scheme.

Lastly, we also notice how, considering the observed mass dependence of final-state correlations~\cite{mass}, it would be interesting to repeate the previous analysis for different flavors and study how hadronization effects are affected by the particle masses. We leave this analysis to a future study.

\section*{Acknowledgements}
We are grateful to A.~Angerami, M.~Clark, S.~Mohapatra and W.~A.~Zajc for the interesting and useful discussions related to the technical aspects of the LHC flow experiments and for the comments about this manuscript. We are also thankful to the whole ATLAS Heavy Ion Group at Columbia University for sharing their computational facilities. The research of AE and MG is supported by U.S. DOE Nuclear Science Grants No. DEFG02-93ER40764.




\begin{thebibliography}{00}


\bibitem{STARAA} 
  J.~Adams {\it et al.}  [STAR Collaboration],
  Nucl.\ Phys.\ A {\bf 757}, 102 (2005)
  [nucl-ex/0501009].
\bibitem{PHENIXAA} 
  K.~Adcox {\it et al.}  [PHENIX Collaboration],
  Nucl.\ Phys.\ A {\bf 757}, 184 (2005)
  [nucl-ex/0410003].
A.~Adare {\it et al.}  [PHENIX Collaboration],
Phys.\ Rev.\ Lett.\  {\bf 105}, 062301 (2010)
[arXiv:1003.5586 [nucl-ex]].



\bibitem{ALICEAA} 
  K.~Aamodt {\it et al.}  [ALICE Collaboration],
  Phys.\ Lett.\ B {\bf 708}, 249 (2012)
  [arXiv:1109.2501 [nucl-ex]].
    K.~Aamodt {\it et al.}  [ALICE Collaboration],
  Phys.\ Rev.\ Lett.\  {\bf 105}, 252302 (2010)
  [arXiv:1011.3914 [nucl-ex]].
    K.~Aamodt {\it et al.}  [ALICE Collaboration],
  Phys.\ Rev.\ Lett.\  {\bf 107}, 032301 (2011)
  [arXiv:1105.3865 [nucl-ex]].
\bibitem{CMSAA} 
  S.~Chatrchyan {\it et al.}  [CMS Collaboration],
  Eur.\ Phys.\ J.\ C {\bf 72}, 2012 (2012)
  [arXiv:1201.3158 [nucl-ex]].
\bibitem{ATLASAA} 
  G.~Aad {\it et al.}  [ATLAS Collaboration],
  Phys.\ Rev.\ C {\bf 86}, 014907 (2012)
  [arXiv:1203.3087 [hep-ex]].
  

  
\bibitem{CMS:2012qk} 
  S.~Chatrchyan {\it et al.}  [CMS Collaboration],
  Phys.\ Lett.\ B {\bf 718}, 795 (2013)
  [arXiv:1210.5482 [nucl-ex]].
\bibitem{Abelev:2012ola} 
  B.~Abelev {\it et al.}  [ALICE Collaboration],
  Phys.\ Lett.\ B {\bf 719}, 29 (2013)
  [arXiv:1212.2001 [nucl-ex]].
\bibitem{ATLASpA} 
  G.~Aad {\it et al.}  [ATLAS Collaboration],
  Phys.\ Rev.\ Lett.\  {\bf 110}, no. 18, 182302 (2013)
  [arXiv:1212.5198 [hep-ex]].
    G.~Aad {\it et al.}  [ATLAS Collaboration],
  Phys.\ Lett.\ B {\bf 725}, 60 (2013)
  [arXiv:1303.2084 [hep-ex]].
\bibitem{Adare:2013piz} 
  A.~Adare {\it et al.}  [PHENIX Collaboration],
  Phys.\ Rev.\ Lett.\  {\bf 111}, no. 21, 212301 (2013)
  [arXiv:1303.1794 [nucl-ex]].
    A.~Dumitru, L.~McLerran and V.~Skokov,
  Phys.\ Lett.\ B {\bf 743}, 134 (2015)
  [arXiv:1410.4844 [hep-ph]].
  
  
  

  
  

\bibitem{BES} 
  L.~Adamczyk {\it et al.}  [STAR Collaboration],
  Phys.\ Rev.\ C {\bf 88}, 014902 (2013)
  [arXiv:1301.2348 [nucl-ex]].



\bibitem{Bozek} 
  P.~Bozek,
  Phys.\ Rev.\ C {\bf 85}, 014911 (2012)
  [arXiv:1112.0915 [hep-ph]].
  P.~Bozek and W.~Broniowski,
  Phys.\ Rev.\ C {\bf 88}, no. 1, 014903 (2013)
  [arXiv:1304.3044 [nucl-th]].
  P.~Bozek and W.~Broniowski,
  Acta Phys.\ Polon.\ B {\bf 45}, no. 7, 1337 (2014)
  [arXiv:1403.6042 [nucl-th]].

\bibitem{ATLASvn} 
  G.~Aad {\it et al.}  [ATLAS Collaboration],
  Phys.\ Rev.\ C {\bf 90}, no. 4, 044906 (2014)
  [arXiv:1409.1792 [hep-ex]].



\bibitem{Raju} 
  K.~Dusling and R.~Venugopalan,
  Phys.\ Rev.\ D {\bf 87}, no. 9, 094034 (2013)
  [arXiv:1302.7018 [hep-ph]].
  A.~Dumitru, K.~Dusling, F.~Gelis, J.~Jalilian-Marian, T.~Lappi and R.~Venugopalan,
  Phys.\ Lett.\ B {\bf 697}, 21 (2011)
  [arXiv:1009.5295 [hep-ph]].
  K.~Dusling and R.~Venugopalan,
  Phys.\ Rev.\ Lett.\  {\bf 108}, 262001 (2012)
  [arXiv:1201.2658 [hep-ph]].
    K.~Dusling and R.~Venugopalan,
  Phys.\ Rev.\ D {\bf 87}, no. 5, 054014 (2013)
  [arXiv:1211.3701 [hep-ph]].
\bibitem{Kovner:2012jm} 
  A.~Kovner and M.~Lublinsky,
  Int.\ J.\ Mod.\ Phys.\ E {\bf 22}, 1330001 (2013)
  [arXiv:1211.1928 [hep-ph]].
\bibitem{McLerran:2014uka} 
  L.~McLerran and V.~V.~Skokov,
  arXiv:1407.2651 [hep-ph].
\bibitem{Levin} 
  E.~Levin and A.~H.~Rezaeian,
  Phys.\ Rev.\ D {\bf 84}, 034031 (2011)
  [arXiv:1105.3275 [hep-ph]].
    E.~Levin and S.~Tapia,
  arXiv:1406.7358 [hep-ph].
\bibitem{Ozonder:2014sra} 
  S.~\"{O}zonder,
  Phys.\ Rev.\ D {\bf 91}, no. 3, 034005 (2015)
  [arXiv:1409.6347 [hep-ph]].
\bibitem{Kovchegov:2012nd} 
  Y.~V.~Kovchegov and D.~E.~Wertepny,
  Nucl.\ Phys.\ A {\bf 906}, 50 (2013)
  [arXiv:1212.1195].
\bibitem{Dumitru}
  A.~Dumitru, A.~V.~Giannini and V.~Skokov,
  A.~Dumitru and A.~V.~Giannini,
  Nucl.\ Phys.\ A {\bf 933}, 212 (2014)
  [arXiv:1406.5781 [hep-ph]].
    A.~Dumitru, F.~Gelis, L.~McLerran and R.~Venugopalan,
  Nucl.\ Phys.\ A {\bf 810}, 91 (2008)
  [arXiv:0804.3858 [hep-ph]].
    A.~Dumitru and V.~Skokov,
  Phys.\ Rev.\ D {\bf 91}, no. 7, 074006 (2015)
  [arXiv:1411.6630 [hep-ph]].
\bibitem{JalilianMarian:1996xn} 
  J.~Jalilian-Marian, A.~Kovner, L.~D.~McLerran and H.~Weigert,
  Phys.\ Rev.\ D {\bf 55}, 5414 (1997)
  [hep-ph/9606337].
\bibitem{Skokov} 
  V.~Skokov,
  Phys.\ Rev.\ D {\bf 91}, no. 5, 054014 (2015)
  [arXiv:1412.5191 [hep-ph]].
    V.~Skokov,
  ``Collectivity in Small Colliding Systems with High Multiplicity,''
  RBRC Workshop, March 2015, http://www.bnl.gov/cscs2015/.


\bibitem{Gyulassy} 
  M.~Gyulassy, P.~Levai, I.~Vitev and T.~S.~Biro,
  Phys.\ Rev.\ D {\bf 90}, no. 5, 054025 (2014)
  [arXiv:1405.7825 [hep-ph]].
    M.~Gyulassy, P.~Levai, I.~Vitev and T.~S.~Bir—,
  Nucl.\ Phys.\ A {\bf } (2014)
  [arXiv:1407.7306 [hep-ph]].



\bibitem{Gunion:1981qs} 
  J.~F.~Gunion and G.~Bertsch,
  Phys.\ Rev.\ D {\bf 25}, 746 (1982).



\bibitem{HIJING} 
  X.~N.~Wang and M.~Gyulassy,
  Phys.\ Rev.\ D {\bf 44}, 3501 (1991).
  
  
\bibitem{Azimov:1984np} 
  Y.~I.~Azimov, Y.~L.~Dokshitzer, V.~A.~Khoze and S.~I.~Troyan,
  Z.\ Phys.\ C {\bf 27}, 65 (1985).

\bibitem{Field:1977fa} 
  R.~D.~Field and R.~P.~Feynman,
  Nucl.\ Phys.\ B {\bf 136}, 1 (1978).


\bibitem{Andersson:1986gw} 
  B.~Andersson, G.~Gustafson and B.~Nilsson-Almqvist,
  Nucl.\ Phys.\ B {\bf 281}, 289 (1987).

  
\bibitem{Sjostrand:1993yb} 
  T.~Sj\"{o}strand,
  Comput.\ Phys.\ Commun.\  {\bf 82}, 74 (1994).


\bibitem{Torrieri:2013aqa} 
  G.~Torrieri,
  Phys.\ Rev.\ C {\bf 89}, no. 2, 024908 (2014)
  [arXiv:1310.3529 [nucl-th]].



\bibitem{mass} 
  S.~Chatrchyan {\it et al.}  [CMS Collaboration],
  Eur.\ Phys.\ J.\ C {\bf 74}, no. 6, 2847 (2014)
  [arXiv:1307.3442 [hep-ex]].
  
    B.~B.~Abelev {\it et al.}  [ALICE Collaboration],
  Phys.\ Lett.\ B {\bf 726}, 164 (2013)
  [arXiv:1307.3237 [nucl-ex]].











\end{thebibliography}


\end{document}